\documentclass[lettersize,journal]{IEEEtran}
\usepackage{graphicx}
\usepackage{textcomp}
\usepackage{xcolor}
\usepackage{subfigure}
\usepackage{comment}
\usepackage{gensymb}
\usepackage{multirow}

\usepackage{amsmath,amsfonts}
\usepackage{algorithmic}
\usepackage{algorithm}
\usepackage{array}
\usepackage{textcomp}
\usepackage{stfloats}
\usepackage{url}
\usepackage{verbatim}
\usepackage{graphicx}
\usepackage{cite}
\begin{document}

\title{Enabling Indoor Multi-Person Tracking With 6G mmWave ISAC Systems}

\author{Chongrui~Wang,~Aimin~Tang,~\IEEEmembership{Senior~Member,~IEEE,}~Fei Gao,~\IEEEmembership{Senior~Member,~IEEE,}~and~Chaojun~Xu
\thanks{Chongrui Wang and Aimin Tang are with Shanghai Jiao Tong University. Fei Gao and Chaojun Xu are with Noika Bell Labs China. Corresponding author: Aimin Tang (email: tangaiming@sjtu.edu.cn).}
}

\maketitle

\begin{abstract}
Integrated sensing and communication (ISAC) has emerged as a key technology for 6G wireless networks. 
In this paper, wireless sensing for the indoor multi-person tracking is explored with 6G mmWave ISAC systems. To limit the sensing overhead, a sparse deployment of sensing reference signals (RS) is applied in the orthogonal frequency-division multiplexing (OFDM) frame, where the channel state information (CSI) at the sensing RS is extracted for the multi-person tracking. To enable a robust tracking of multiple persons in a complex indoor environment, three key mechanisms are proposed: 1) a modified moving target indicator (MTI) scheme is proposed to remove the static environmental clutter under a sparse RS time spacing; 2) an effective target identification mechanism is developed to exclude false target points; 3) the Kalman filter with a penalty association algorithm is designed to associate the detected points with the right tracks, especially handling the crossover case of two tracks. With the above mechanisms, multiple persons can be effectively tracked with an extremely low sensing overhead. An mmWave bistatic ISAC prototype system at 26 GHz with 500 MHz bandwidth has been developed to validate our design, where the overhead of the sensing RS is less than 0.005\%. Experimental results demonstrate that our proposed design achieves a median position error of 12 cm for multi-person tracking with path-crossing in the indoor environment with a single receiver.
\end{abstract}

\section{Introduction}

Integrated sensing and communication (ISAC) has been envisioned as a cornerstone for 6G systems \cite{liu2022integrated}. 
The indoor localization and tracking are important applications for ISAC, which have attracted great attention in recent years. 

Due to the widespread deployment of WiFi systems in indoor scenarios, WiFi-based localization/tracking has been widely investigated. 
In WiFi systems, the channel state information (CSI) can be extracted from the preamble of each frame for sensing. So far, the indoor single-target localization and tracking have been widely studied via commercial WiFi systems  \cite{WiFiSingle,Widar,WiDance,WiDir}.
When multiple persons perform actions simultaneously, their activities will generate an aggregated influence on the CSI, which extraordinarily increases the difficulties of tracking. A few studies have been carried out to address this problem. In \cite{trackingFromOneSide}, a framework for multi-person tracking is introduced using only the signal magnitude from a single transmitter and a receiver array placed on one side of the area by jointly estimating AoA and motion-induced parameters. The system in \cite{MultiTrack} leverages multiple WiFi links and splices together all available channels in the 5 GHz band to enable multi-user tracking in indoor environments.

\begin{figure*}
    \centering
    \includegraphics[width=0.8\linewidth]{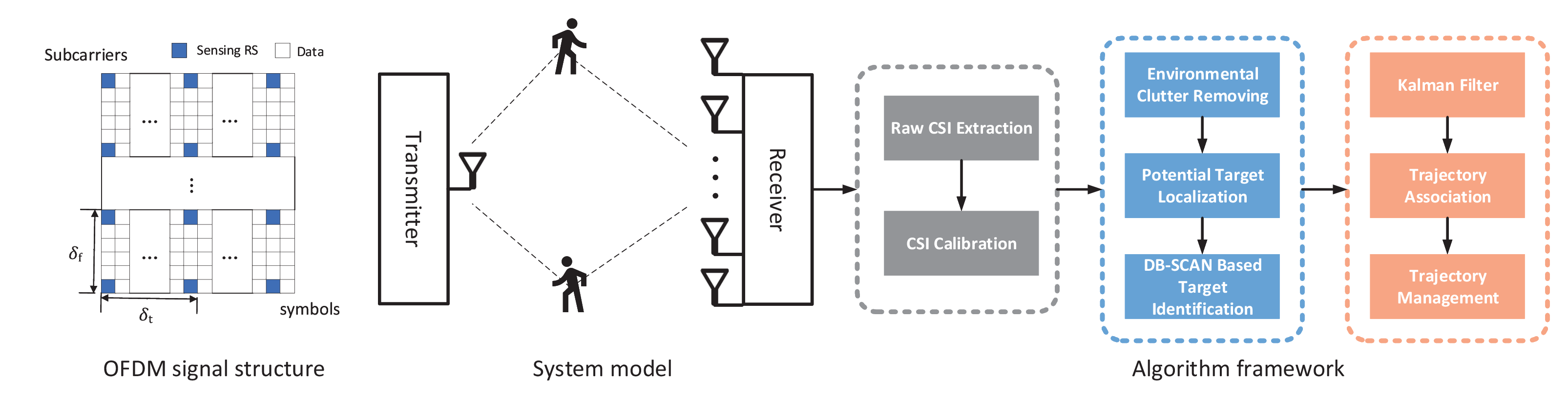}
    \vspace{-1em}
    \caption{Signal structure, system model, and algorithm framework of our proposed design.}
    \vspace{-1em}
    \label{fig:framework}
\end{figure*}

Despite the widely studied WiFi-based solutions, only a few explorations have been carried out based on 5G/6G ISAC systems, where the CSI is extracted from the reference signals (RS) for sensing. In \cite{AndrewISACSensing}, a robust ISAC framework is established for precise single-person tracking by extracting refined Doppler-delay-angle features. In \cite{li2023contact}, multi-person tracking is achieved by the deployment of multiple receivers in a distributed MIMO setup. In \cite{li2023integrating}, multi-person localization is demonstrated with mmWave ISAC prototype systems in the indoor environment. In \cite{luo2025passive}, a 26 GHz mmWave passive bistatic ISAC prototype is developed, which extracts multipath propagation parameters and adopts power-weighted clustering to realize joint localization of static scatterers and moving targets with sub-meter level accuracy in outdoor open square scenarios.
In \cite{liu2023integrated}, a field trial of an ISAC system is built, which utilizes standard base stations without hardware modification and employs the standardized remote-interference-management (RIM) signal for sensing. 
However, it does not address the critical challenges posed by complicated environmental clutter and the multi-target path-crossing problem. Thus, the potential of utilizing cellular ISAC systems for indoor multi-person tracking is still underexplored. 

In this paper, the indoor multi-person tracking is explored based on 6G bistatic mmWave ISAC systems. 
We develop a complete sensing/tracking pipeline that effectively addresses clutter interference, effective target detection, and cross-target association errors. Specifically, three key mechanisms are proposed. 1) A modified moving target indicator (MTI) algorithm is developed that can remove static environmental clutter while preserving weak dynamic echoes caused by human motion. 2) A CIR-guided-search-pruning-2D-MUSIC-based localization method is developed to jointly estimate delay and AoA with super-resolution capability, enabling precise target positioning even under sparse RS deployment. A density-based spatial clustering of applications with noise (DBSCAN) method is adopted to eliminate falsely detected targets in the cluttered indoor environment. 3) A constant velocity (CV)-based Kalman filtering (KF) integrated with a penalty association algorithm is proposed, which explicitly addresses the path-crossing problem, where two targets' trajectories intersect. 

An mmWave OFDM ISAC prototype system at 26 GHz is developed to validate the proposed design. Experimental results demonstrate that a median position error of 12 cm is achieved for multi-person tracking.
Compared to the existing studies, our proposed approach achieves the above high tracking accuracy with a setup of a single receiver in the multi-person path-crossing scenario with negligible sensing overhead (less than 0.5‱).


\section{System and Signal Model}
The system model is shown in Fig. \ref{fig:framework}. The ISAC system comprises a single-antenna transmitter and a sensing receiver equipped with a uniform linear array (ULA) with $P$ antennas to enable AoA estimation. 
The sensing RSs are embedded into the data frame in a scattered way, as shown in Fig. \ref{fig:framework}. 
While the data payload is intended for communication users and is private, the sensing RSs are designed as public, which means that the sensing receiver, no matter whether it is the intended communication receiver, can extract CSI for wireless sensing based on the public/known sensing RSs.

The sensing RS pattern is carefully designed for high sensing performance and low overhead. The RSs are scattered equally throughout the OFDM signal to suppress sensing sidelobes \cite{zhao2023reference}. The frequency- and time-spacing for RS are denoted as $\delta_\text{f}$ and $\delta_\text{t}$, respectively. Therefore, the overhead is only $\frac{1}{\delta_\text{f}\delta_\text{t}}$. The RSs are sparsely placed with large frequency and time intervals. In the frequency domain, sensing RSs span the entire bandwidth to harvest the maximum delay resolution. However, the selection of $\delta_\text{f}$ should keep range detection unambiguous for indoor scenarios. To achieve an extremely low overhead, the time-spacing $\delta_\text{t}$ can be designed in a very sparse way. However, this will result in severe Doppler ambiguity, making the Doppler information unusable, which poses an additional challenge for the tracking method. 

The element of CSI matrix $\mathbf{H}_{p}$ on the $n \delta_\text{f}$-th subcarrier of the $m \delta_\text{t}$-th symbol can be modeled as
\begin{equation}
{\mathbf{H}}_{p}[m,n]=\sum_{l=1}^L{\boldsymbol{a}_{\boldsymbol{l}}[p]b_{l}}e^{-j2\pi n\delta_\text{f} \Delta_f \tau _l}e^{j2\pi m \delta_\text{t}T f_{\text{D},l}},
\label{eq:oriCSI}
\end{equation}
where $L$ is the total number of propagation paths from the transmitter to the sensing receiver; $\boldsymbol{a}_{\boldsymbol{l}}$ is the steering vector for the $l$-th path and $\boldsymbol{a}_{\boldsymbol{l}}=[1,e^{j2\pi\frac{d}{\lambda}\sin\phi_l},...,e^{j2\pi(P-1)\frac{d}{\lambda}\sin\phi_l}]$, where $d$ is the distance interval of adjacent antenna, $\lambda$ is the wavelength of the signal, and $\phi_l$ is the AoA of the $l$-th path; 
$\tau_l$ and $f_{\text{D},l}$ denote the delay and Doppler shift of the $l$-th path, respectively; $\Delta f$ and $T$ denote the OFDM subcarrier spacing and symbol duration, respectively; 
$b_{l}$ is the complex attenuation of the $l$-th path. 

The CSI $\mathbf{H}_{p}[m,n]$ for sensing is estimated from the sensing RS using the least squares (LS) algorithm. We assume that impairments in bistatic ISAC systems, including sampling time offset (STO), carrier frequency offset (CFO), and random phase noise, have already been compensated for by employing our previously proposed reference-path-aided calibration scheme described in \cite{luo2025experimental}, which can achieve a sub-nanosecond synchronization performance.

\section{Algorithm Design}
\subsection{Framework Design}

The overall system framework is illustrated in Fig.~\ref{fig:framework}. After the raw CSI is extracted, our proposed reference-path-aided calibration \cite{luo2025experimental} is applied to eliminate the STO, CFO, and random phase shift. Next, a modified MTI Filter is developed to suppress static environmental clutter. Subsequently, a localization-tracking pipeline is executed. The localization stage extracts the delay and angle of each significant reflection path to compute the real-time positions of potential targets, followed by a DBSCAN-based method to eliminate false detections. The tracking stage then employs a multi-target algorithm, combining a Kalman filter for state estimation with a penalty Hungarian algorithm for trajectory association. This pipeline robustly reconstructs the continuous trajectories of multiple targets, effectively managing temporary occlusions and the emergence of new targets.


\subsection{Environmental Clutter Removing} 
Before performing target localization, it is essential to suppress the environmental clutter. In indoor environments, rich multi-path reflections often dominate the received signal power, causing numerous false targets if processed directly. 
Hence, background clutter removal is a necessary preprocessing step for reliable target detection.



Basically, the CSI $\mathbf{H}_p$ can be decomposed into a static part and a dynamic part as $\mathbf{H}_p = \mathbf{H}_{p,\text{static}} + \mathbf{H}_{p,\text{dynamic}}$.
The $\mathbf{H}_{p,\text{static}}$ is environmental clutter, remains unchanged across time, and should be eliminated. The typical low-complexity clutter removal technique is MTI, which suppresses static clutter by subtracting consecutive channel observations, i.e., ${\mathbf{H}}_{p}[m] - {\mathbf{H}}_{p}[m-1]$. However, this approach fails to detect slow-moving humans, as their dynamic channel component $\mathbf{H}_{p,\text{dynamic}}$ remains nearly unchanged between observations, causing it to be canceled along with the clutter. This occurs, for instance, when a person is stationary but breathing.

To resolve this problem, we propose a modified MTI algorithm based on temporal averaging, which is given by:
\begin{equation}
\begin{split}
    &\mathbf{H}'_{p}[m,n]
    = \mathbf{H}_{p}[m,n]
    - \frac{1}{K}\sum_{k=1}^{K}\mathbf{H}_{p}[m-k,n]\\
    &=
    \mathbf{H}_{p,\text{dynamic}}[m,n]
    - \frac{1}{K}\sum_{k=1}^{K}\mathbf{H}_{p,\text{dynamic}}[m-k,n].
\end{split}
\end{equation}
We assume that the delay of each path remains approximately constant over the considered time window. This is reasonable because the target displacement between two consecutive RS symbols is on the order of millimeters, while the corresponding phase contribution from delay variation is smaller by far than the Doppler term. Let $\Delta\Phi=2\pi f_D\Delta T$ denote the phase increment between two consecutive RS symbols, where $f_D$ is the Doppler frequency $f_D=\frac{2v}{\lambda}$. The magnitude of the average term (the subtractor) can be derived as the magnitude of a geometric series sum:
\begin{equation}
\begin{split}
&\left|
\frac{1}{K}\sum_{k=1}^{K}\mathbf{H}_{\text{dynamic}}[m-k]
\right|
=
\left|
\frac{1}{K}\sum_{k=1}^{K}Ae^{-jk\Delta\Phi}
\right| \\
&=
A\cdot \frac{\left| \sin{ (K\Delta\Phi / 2) } \right| }
{K \left| \sin{(\Delta\Phi / 2 )} \right|}.
\label{eq:MTI2}
\end{split}
\end{equation}
Here, we ignore the indices $p$ and $n$ for simplicity. Eq. (\ref{eq:MTI2}) is expected to approach 0 for a large $K$ so that we have $\mathbf{H}'_{p}[m,n] \approx \mathbf{H}_{p,\text{dynamic}}[m,n]$. 


\subsection{Effective Target Localization and Identification}


To enable multi-person localization and tracking with a single sensing receiver, a super-resolution algorithm is essential. To this end, we leverage the available degrees of freedom in both range and angle by employing a high-resolution 2D-MUSIC algorithm for joint estimation. 
The first step for 2D-MUSIC is constructing the covariance matrix. 
Given $\mathbf{H}'_{p}[t,n]$ in a radar CPI with $T$ snapshots, we firstly form a stacked data vector for each snapshot $t$ by vectorizing the antenna--frequency matrix into a column vector $\mathbf{x}_t \;=\; \operatorname{vec}\!\big( [\,\mathbf{H}'_{p}[t,n]\,]_{p=1..P,\;n=0..N-1} \big)
\in \mathbb{C}^{PN \times 1}.$
For $T$ snapshots, the covariance matrix can be estimated by $\mathbf{R} \;=\; \frac{1}{T}\sum_{t=1}^{T} \mathbf{x}_t \mathbf{x}_t^H
\in \mathbb{C}^{PN\times PN}.$
Singular value decomposition (SVD) is further performed on $\mathbf{R}$ to decompose the signal subspace and noise subspace as
\begin{equation}
    \mathbf{R} = \mathbf{E}\,\boldsymbol{\Lambda}\,\mathbf{E}^H
= \mathbf{E}_s\boldsymbol{\Lambda}_s\mathbf{E}_s^H
+ \mathbf{E}_n\boldsymbol{\Lambda}_n\mathbf{E}_n^H,
\end{equation}
where $\mathbf{E}_s$ and $\mathbf{E}_n$ are the signal and noise subspace eigenvector matrices, respectively, and $\boldsymbol{\Lambda}_s,\boldsymbol{\Lambda}_n$ are the corresponding diagonal eigenvalue matrices sorted in descending order. 
The 2D-MUSIC pseudo-spectrum is then evaluated by testing candidate pairs $(\tau,\varphi)$ and measuring their orthogonality to the noise subspace:
\begin{equation}
P_{\text{MUSIC}}(\tau,\varphi)
\;=\;
\frac{1}{\mathbf{s}(\tau,\varphi)^H \,\mathbf{E}_n\mathbf{E}_n^H\, \mathbf{s}(\tau,\varphi)},
\label{eq:music}
\end{equation}
where $\mathbf{s}(\tau,\varphi)$ is the joint delay–angle steering vector. 
Peaks of $P_{\text{MUSIC}}(\tau,\varphi)$ indicate the estimated joint delay–angle parameters $\{\hat{\tau}_l,\hat{\varphi}_l\}$.

To reduce computational complexity, we introduce a \textbf{CIR-guided search pruning} strategy.
Since the CIR already provides rough delay information, we first identify potential delay peaks $\{\tilde{\tau}_l\}$ from the CIR obtained by the inverse fast Fourier transform (IFFT) along subcarriers:
\begin{equation}
Q_p[m,u] = \frac{1}{N} \sum_{n=0}^{N-1} H'_{p}[m,n] e^{j2\pi n u / N_{\mathrm{ifft}}}.
\end{equation}
Then, instead of performing an exhaustive 2D search over the entire delay–angle grid, the MUSIC spectrum is evaluated only around each $\tilde{\tau}_l$ and its neighboring delay bins.

The reduction of computational complexity is analyzed as follows. Given 2D-MUSIC dimension $D=P\times N$, full-search delay bins $G_\tau$, full-search AoA bins $G_\varphi$, and pruned-search delay bins via CIR-guided searching $G_{\tau,\text{pruned}}$, the overall computational complexity of the conventional full-search 2D-MUSIC is given by: $\mathcal{O}(T_sD^2+D^3+G_\tau\cdot G_\varphi\cdot D^2)$, while the complexity of the CIR-guided search-pruned 2D-MUSIC becomes: $\mathcal{O}\left(T_sD^2+D^3+N_{\mathrm{ifft}}\log{N_{\mathrm{ifft}}}+G_{\tau,\text{pruned}}\cdot G_\varphi\cdot D^2\right)$. Since the computational complexity is dominated by the two-dimensional spectrum searching, the pruning approach significantly reduces computation (although adding additional complexity for CIR calculation), while preserving accuracy, as the search space is restricted to the physically meaningful regions indicated by the CIR peaks.

\begin{figure}[t]
    \centering
    \includegraphics[width=0.6\linewidth]{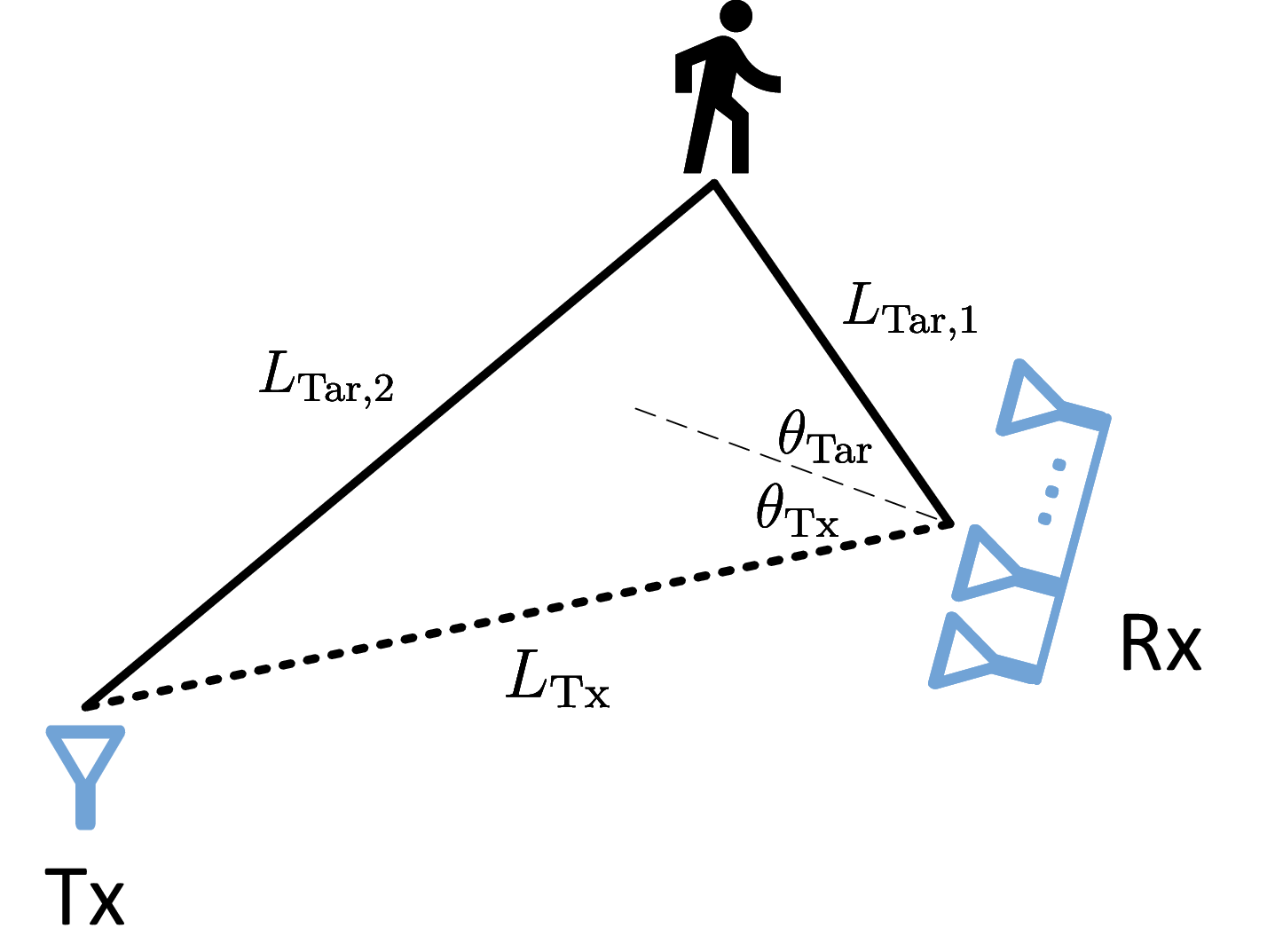}
    \vspace{-1em}
    \caption{Geometry model for target localization under a single sensing receiver.}
    \vspace{-1em}
    \label{fig:geomodel}
\end{figure}

For each detected peak $(\hat{\tau}_l, \hat{\varphi}_l)$ on the 2D-MUSIC spectrum, the corresponding target follows a specific geometry relationship under the bistatic sensing architecture, as shown in Fig. \ref{fig:geomodel}. 
More specifically, the geometry follows:
\begin{equation}
\label{eq:geometry1}
\begin{cases}
    {L_{\text{Tx}}}^2 + {L_{\text{Tar,1}}}^2 - 2 {L_{\text{Tx}}}{L_{\text{Tar,1}}} \cos{(\theta _{\text{Tar}} + \theta _{\text{Tx}})} = {L_{\text{Tar,2}}}^2 \\
    {L_{\text{Tar,1}}} + {L_{\text{Tar,2}}} = {L_{\text{Tar}}} = c \hat{\tau}_l \\
    \theta_{\text{Tar}} = \pi - \hat{\varphi}_l
\end{cases}
\end{equation}
where $\theta _{\text{Tar}}$ is calculated based on the normal angle of the sensing receiver and the estimated AoA, and $\theta _{\text{Tx}}$ is prior information, i.e., AoA of the transmitter. Then, we can derive from the equation \eqref{eq:geometry1} that
\begin{equation}
{L_{\text{Tar,1}}} = \frac{{L_{\text{Tar}}}^2 - {L_{\text{Tx}}} ^2}{2{L_{\text{Tar}}} - 2{L_{\text{Tx}}} \cos{(\theta _{\text{Tar}} + \theta _{\text{Tx}})}}
\end{equation}
The target position $(x_l, y_l)$ is derived as:
\begin{equation}
    (x_l, y_l) = (x_{\text{Rx}}, y_{\text{Rx}}) + {L_{\text{Tar,1}}}(\cos \hat{\varphi}_l, \sin \hat{\varphi}_l)
\end{equation}

After environmental clutter removal, our experimental study shows that estimated potential targets still contain a few false reflections, referred to as \textit{fake targets}. These fake targets may originate from minor environmental disturbances, such as small mechanical vibrations or subtle object movements. Such reflections typically last for only a few sensing frames and are hard to remove in the modified MTI algorithm. These transient echoes would manifest as isolated target points, introducing severe noise in the subsequent tracking stage. To further eliminate fake targets, the DBSCAN algorithm is further employed. Each cluster identified by DBSCAN corresponds to one effective physical target. The centroid of each cluster is then calculated as the representative position.


\subsection{Multi-Target Tracking}


After DBSCAN-based target identification, each target is modeled under a constant-velocity (CV) assumption, which reflects the typical motion pattern of humans walking in indoor environments. Even when multiple people cross paths, their motion directions generally remain unchanged for a short duration. This observation motivates the use of the CV-based Kalman filter model to suppress false associations caused by trajectory intersection.

The CV-based Kalman filter predicts the position of each target at every frame according to the CV motion model. 
These predicted positions are then used in the subsequent trajectory association stage, effectively preventing false associations when multiple targets cross paths, since the prediction constrains each trajectory to continue along its prior motion direction. 
After trajectory association, the Kalman filter will incorporate the newly observed positions to correct the predicted states, thereby smoothing short-term motion fluctuations and maintaining continuous, physically consistent trajectories. 

\begin{figure}[t]
    \centering
    \includegraphics[width=0.6\linewidth]{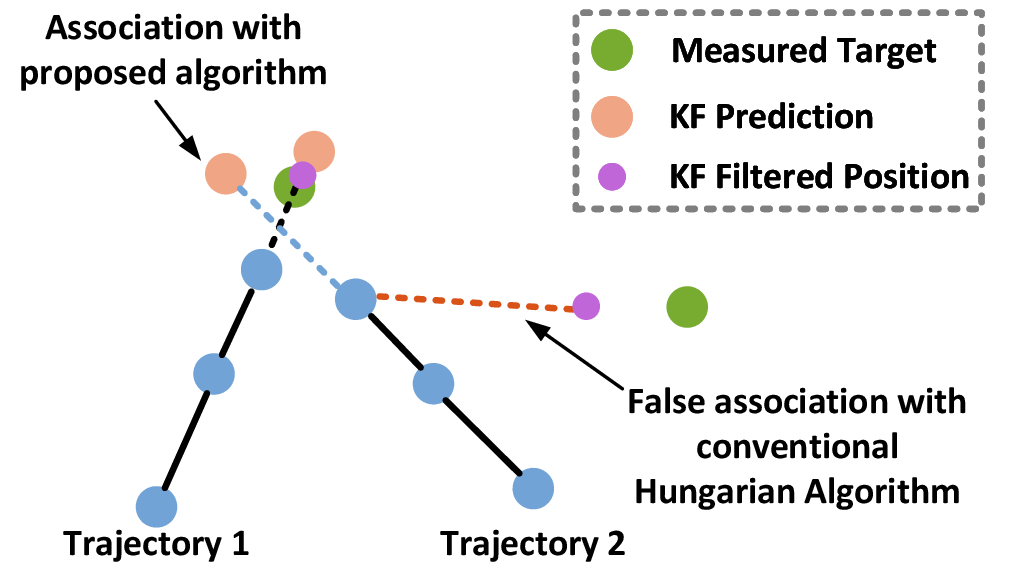}
    \vspace{-1em}
    \caption{Example of trajectory association with path crossing under different algorithms.}
    \label{fig:HungarianMismatch}
    \vspace{-1em}
\end{figure}

For multi-target tracking, trajectory association is the key to ensuring that each detection is correctly matched to its corresponding trajectory. For each frame, assume that there are $N$ newly detected targets, denoted as $\{P_j\}_{j=1}^{N}$, and there are $M$ current trajectories with $\{\hat{P}_i\}_{i=1}^{M}$ predicted points from Kalman Filter.  
The association cost between trajectory $i$ and detection $j$ is defined as the Euclidean distance, i.e., $c_{ij} = \| \hat{P}_i - P_j \|$.
The trajectory association problem is typically addressed by the conventional Hungarian algorithm\cite{kuhn1955hungarian}, which aims to minimize the total cost by performing a \textit{full assignment} between all trajectories and detections. 
This global matching mechanism of the conventional Hungarian algorithm has a critical problem in practical applications when a real target is missed in the detection, while a new target appears at the same time (may be either a falsely detected target or a new real target). Target missing can happen in two typical cases: 1) the crossover of two trajectories leads to unresolved targets; 2) one target temporarily disappears due to occlusion. Under these critical cases, the conventional Hungarian algorithm may force an incorrect matching, as shown in Fig. \ref{fig:HungarianMismatch}. 
To address this issue, a \textbf{penalty association algorithm} is proposed by introducing a distance penalty. If the Euclidean distance between a predicted position and a detected point exceeds a threshold $d_{\text{th}}$, a large penalty $d_\text{p}$ is assigned. Therefore, the cost $c_{ij}$ between $\hat{P}_i$ and $P_j$ is designed as:
\begin{equation}
c_{ij} =
\begin{cases}
\| \hat{P}_i - P_j \|, & \text{if } \| \hat{P}_i - P_j \| \le d_{\text{th}}, \\
d_\text{p}, & \text{otherwise.}
\end{cases}
\end{equation}
The optimal assignment is resolved by first solving
\begin{equation}\label{prob:associ}
\min_{\delta_{ij}} \sum_{i=1}^{M} \sum_{j=1}^{N} c_{ij} \delta_{ij}
\end{equation}
\begin{equation*}
\text{s.t. } 
\delta_{ij} \in \{0,1\},
\end{equation*}
\begin{equation*}
    \begin{split}
        \forall{i}, \sum_j \delta_{ij} = 1,\text{ if } M\leq N;
        \quad
        \forall{j}, \sum_i \delta_{ij} = 1,\text{ if } M > N.
    \end{split}
\end{equation*}
The Hungarian algorithm can now be applied to solve problem (\ref{prob:associ}). If the number of trajectories is very small, the solution to problem (\ref{prob:associ}) can also be found by exhaustive searching. With the solution, we then set $\delta_{ij} = 0$ for those with penalty cost.
If a trajectory is not associated with any detected target, the predicted point is used in this frame. 
This modification effectively prevents the algorithm from making forced associations when a trajectory temporarily loses its target and a new target emerges.
This adaptive strategy significantly improves robustness in dynamic multi-person scenarios.

After association, trajectory states are continuously updated and evaluated.
A new trajectory is initialized if a new detection persists for several consecutive frames, while trajectories without updates for a predefined duration are terminated.
Short-lived or unstable trajectories are regarded as false tracks and discarded.

\section{Experimental Evaluation}
  
\begin{figure}[t]
    \begin{minipage}{0.42\linewidth}
        \centering
        \includegraphics[width=1\linewidth]{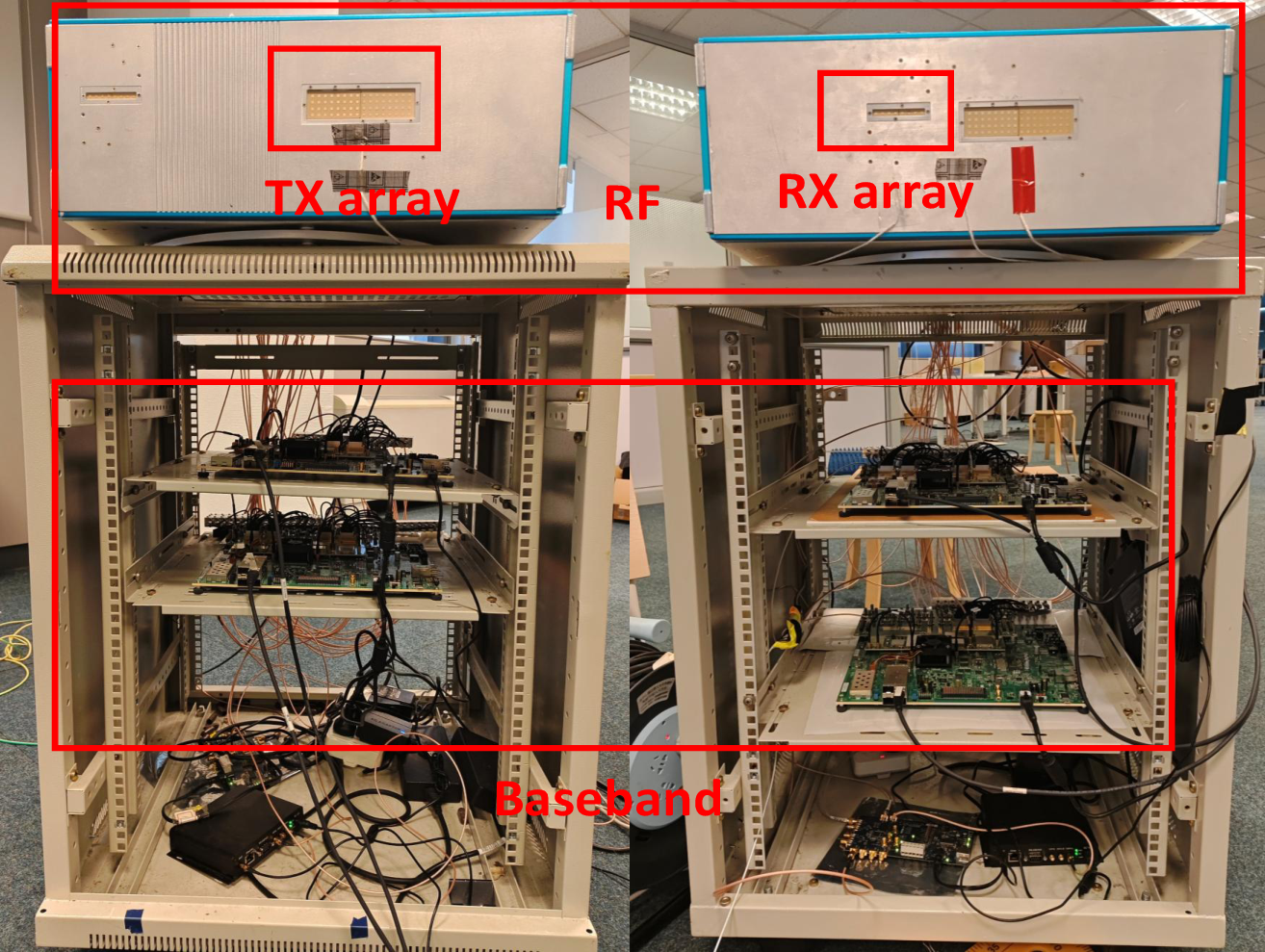}\\
        \caption{Our developed mmWave ISAC prototype system.}
        \label{fig:ESsetup}
    \end{minipage}
    \hfill
    \begin{minipage}{0.54\linewidth}
        \centering
        \vspace{-1em}
        \includegraphics[width=1\linewidth]{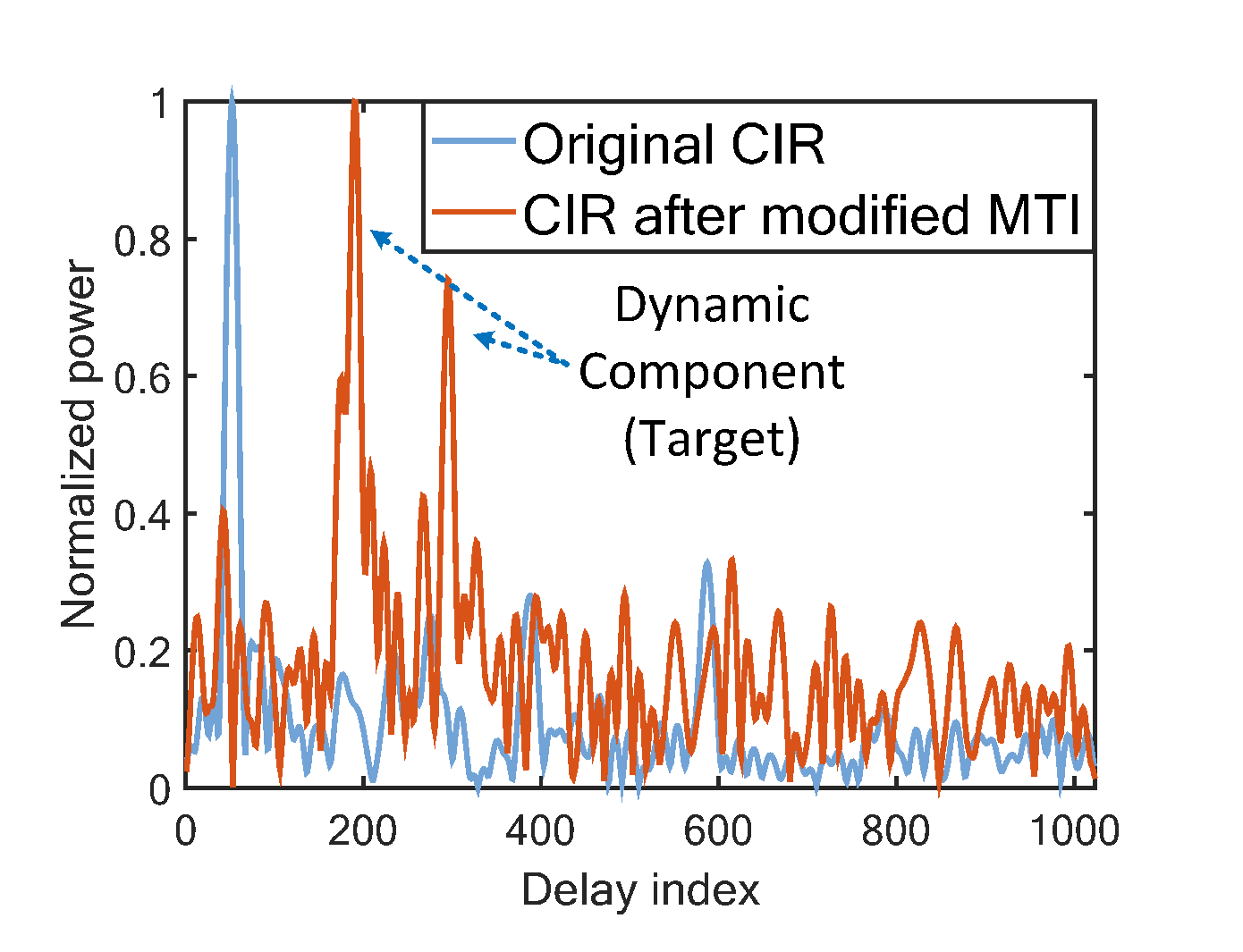}\\
        \vspace{-1.5em}
        \caption{CIR before and after the modified MTI algorithm.}
        \label{fig:exp_mti}
    \end{minipage}
    \vspace{-1em}
\end{figure}

An mmWave OFDM ISAC prototype system at 26 GHz with 500 MHz bandwidth is developed to validate the effectiveness of the proposed design, as shown in Fig. \ref{fig:ESsetup}. Due to the page limitation, the detailed configuration of the prototype can be referred to \cite{11477778}.
The ISAC frame structure deploys sensing RS with frequency spacing $\delta_\text{f}=24$ subcarriers, resulting in $N=76$ RS subcarriers spaced at $6.48$~MHz intervals. Each OFDM block has a duration of $4$~ms and contains $864$ symbols. The sensing RS is inserted in every OFDM block, corresponding to a time spacing of $\delta_\text{t}=864$ symbols ($4$~ms). Thus, the resulting RS overhead is $\frac{1}{24 \times 864} = 0.0048\%$.
This minimal overhead is negligible for communication systems. As for the penalty association algorithm, the distance threshold $d_\text{th}$ is set to 1 m to balance tolerance for localization errors (typically 10–30 cm) and avoidance of mismatches, especially during trajectory crossing. The penalty $d_\text{p}$ is set to 1000 m, much larger than $d_\text{th}$ and the maximum feasible distance in the sensing area, to strongly discourage invalid forced associations in the optimization.


\begin{figure*}[t]
    \centering
    \subfigure[2-person non-crossing tracking]{
    \includegraphics[width=0.22\linewidth]{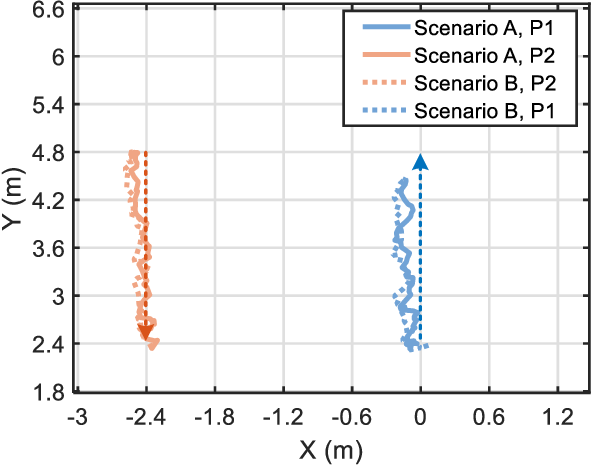}
    \label{fig:exp_dual_noncross}\vspace{-10mm}
    }
    \hspace{-2mm}
    \subfigure[2-person crossing tracking]{
    \includegraphics[width=0.22\linewidth]{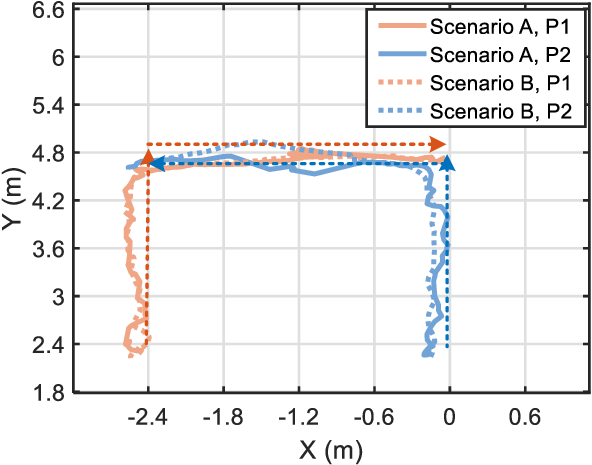}
    \label{fig:exp_cross}\vspace{-10mm}
    }
    \hspace{-2mm}
    \subfigure[3-person tracking]{
    \includegraphics[width=0.22\linewidth]{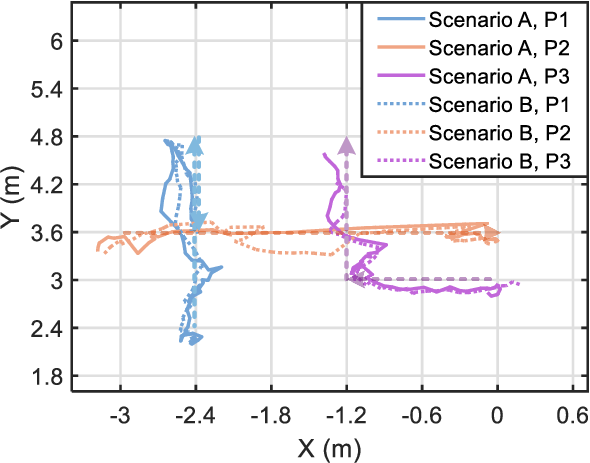}
    \label{fig:exp_3people}\vspace{-10mm}
    }
    \hspace{-2mm}
    \subfigure[CDF for all cases]{
    \includegraphics[width=0.24\linewidth]{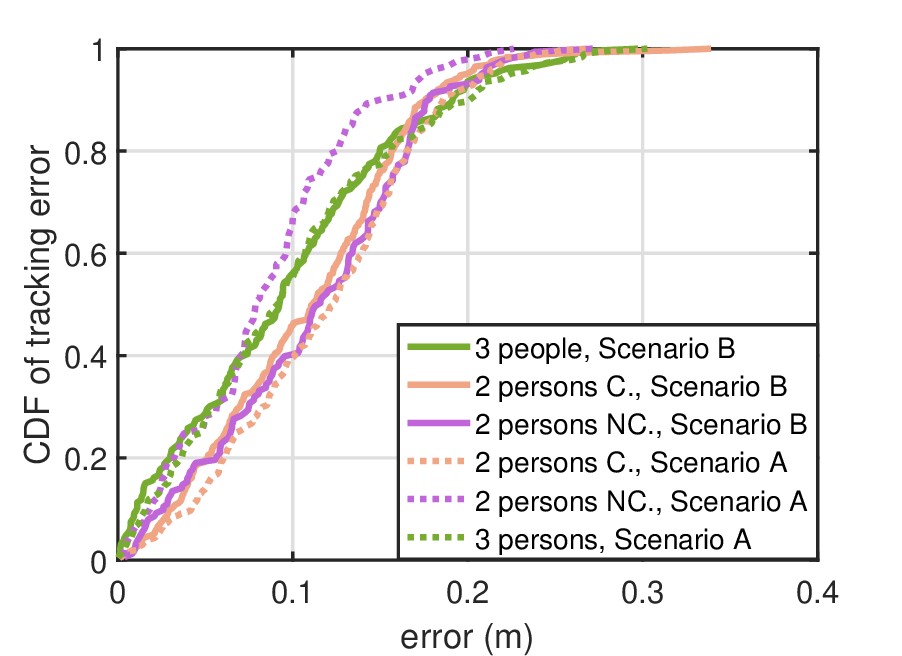}
    \label{fig:cdf}\vspace{-10mm}
    }
    \caption{Experimental evaluation of the proposed localization and tracking framework.}
    \label{fig:exp_results}
    \vspace{-1em}
\end{figure*}

We first evaluate the performance of the proposed modified MTI algorithm in clutter removal. Fig.~\ref{fig:exp_mti} compares the CIR before and after modified MTI filtering, where K is empirically chosen as 50. The original CIR contains strong static reflections from walls and furniture, which dominate the dynamic echoes. After applying the modified MTI, static clutter is effectively suppressed, and the dynamic reflections caused by human motion are clearly preserved. This shows the effectiveness of the proposed algorithm.

We then carry out a series of experiments to verify the effectiveness of the proposed localization and tracking framework under different scenarios.
All experiments are performed in two different indoor laboratory environments, both containing rich environmental clutter.
Three representative experimental cases were tested in both scenarios:
(a) 2-person non-crossing tracking,
(b) 2-person path-crossing tracking, and
(c) 3-person path-crossing tracking.
For each case, both qualitative trajectory reconstruction and quantitative accuracy statistics are analyzed.
In the 2-person non-crossing tracking scenario, two subjects walk simultaneously in the sensing area along parallel paths, as shown in Fig.~\ref{fig:exp_dual_noncross}, where the two reconstructed trajectories remain clearly separated. In the 2-person path-crossing tracking scenario, two subjects are instructed to walk toward each other and cross at approximately the center of the sensing area. This is a challenging scenario where the targets' instantaneous range-angle positions overlap, easily causing identity swapping in conventional tracking algorithms. As shown in Fig.~\ref{fig:exp_cross}, even when the two subjects intersect, the trajectories remain continuous and identity-consistent. In the 3-person path-crossing tracking scenario, three subjects are instructed to walk paths that cross. As shown in Fig.~\ref{fig:exp_3people}, the trajectories remain continuous and identity-consistent, showing the robustness of the proposed pipeline. The CDF comparisons of these experiments are plotted in Fig.~\ref{fig:cdf}. It can be seen that all cases achieve similar performance; the median tracking error is less than 12~cm, with 90\% of the samples remaining below 20~cm.

\section{Conclusion}
In this paper, a robust localization and tracking framework was proposed for multi-target sensing with 6G mmWave ISAC systems. A modified MTI filter was first applied to eliminate environmental clutter. Then, a high-resolution 2D-MUSIC-based localization scheme combined with a DBSCAN clustering algorithm was developed to effectively remove short-lived fake targets caused by minor environmental perturbations. For continuous multi-target tracking, a Kalman filter with a penalty association strategy was introduced, effectively mitigating false associations during target trajectory crossings and temporary occlusions. Experimental results with a 26~GHz ISAC prototype demonstrated that the proposed method achieves accurate, stable, and continuous trajectory reconstruction under low RS overhead and complex indoor environments, showing strong potential for future intelligent sensing and human motion monitoring applications.

\bibliographystyle{IEEEtran}
\bibliography{IEEEabrvRef}

\end{document}